# Quality control and quality assurance evaluation of ALFE2, a large-dynamic-range front-end ASIC developed for the ATLAS Liquid Argon Calorimeter high-luminosity LHC upgrade

---

**E. Buschmann,[a] G. Carini,[b] G. Chatzianastasiou,[a] H. Chen,[a] Y. Chen,[c] M. Dabrowski,[b,1] G. Deptuch,[b] L. Duflot,[d] M. Feo,[a] J. Kierstead,[b] T. Liu,[a,*] H. Ma,[a] D. Matakias,[a,2] N. Morange,[d] M. Oliveira,[a] S. Rescia,[b] E. Rossi,[a] S. Tang,[a] M. Tamari,[a] H. Xu[a]**

[a] *Physics Department, Brookhaven National Laboratory*
*Upton, NY 11973, United States of America*

[b] *Instrumentation Department, Brookhaven National Laboratory*
*Upton, NY 11973, United States of America*

[c] *Department of Physics and Astronomy, Stony Brook University*
*Stony Brook, NY 11794, United States of America*

[d] *IJCLab, Université Paris-Saclay, CNRS/IN2P3*
*Orsay, 91405, France*

[1.] *Now at ESAT-ADVISE, KU Leuven, Leuven, 3000, Belgium*

[2.] *Now at CERN, 1211 Geneva 23, Switzerland*

*E-mail*: tliu5@bnl.gov

ABSTRACT: ALFE2 is a front-end ASIC developed for the ATLAS Liquid Argon (LAr) Calorimeter upgrade during the High-Luminosity Large Hadron Collider (HL-LHC) phase. ALFE2 comprises four preamplifier/shaper channels, each providing two distinct gain outputs to cover a 16-bit dynamic range. A robotic system has been developed for the automatic quality control test of ALFE2, and over 10% of the 80,000 chips have been evaluated by September 2025. The evaluation has allowed us to establish grading criteria. Using these criteria, a yield of over 85% was achieved in the evaluation tests, and these criteria are now being applied to the ongoing full-production QC. Irradiation tests were also performed for the quality assurance of ALFE2. No significant performance degradation was observed during the total-ionizing-dose (TID) test. Based on the single-event effect (SEE) test results, an error rate of fewer than 4.6 single-event upsets (SEUs) per day is extrapolated for the entire ATLAS LAr Calorimeter during HL-LHC operation.

## Contents

## 1. Introduction

The High-Luminosity Large Hadron Collider (HL-LHC) upgrade, targeting an instantaneous luminosity up to $7.5\times10^{34}$ $cm^{-2}s^{-1}$, is scheduled to commence data taking in 2030. During this upgrade, the readout electronics of the ATLAS Liquid Argon (LAr) Calorimeter will be replaced to accommodate higher data rates and radiation levels [1][2]. The ATLAS LAr front-end (ALFE) chip, a large-dynamic-range, low-noise, high-linearity preamplifier/shaper ASIC, is developed for the HL-LHC upgrade [3][4][5][6][7]. ALFE2 is the final production version of ALFE.

A block diagram of ALFE2 is shown in Figure 1(a). ALFE2 integrates quad-channel transimpedance PreAmplifiers (PAs) and CR-(RC)$^2$ shapers and is required to cover a 16-bit dynamic range. To achieve this large dynamic range, each channel provides two differential outputs with different gains: low gain (LG) and high gain (HG). In addition, both outputs can be read out in parallel. Each HG or LG output includes a CR-RC shaping stage and an RC shaping stage. To ensure compatibility with the current Level-0 trigger system [8], ALFE2 includes a differential trigger-sum (TS) output, which is based on the low-gain outputs of the preamplifiers. CR-RC shaping is implemented in the TS output because anti-aliasing is performed later in the trigger chain.

The ATLAS LAr Calorimeter front-end electronics system is located outside the liquid argon cryostat and operates at room temperature. Transmission lines couple detector cells to the preamplifiers. These preamplifiers are required to have a controlled input impedance for proper termination. Two transmission-line impedances are adopted: 50 Ω for pre-sampler and front layers and 25 Ω for middle and back layers [9]. The capacitances (330 pF for the front layer and 1.5 nF for the middle layer) of the typical calorimeter cells in the barrel area are chosen as the reference load for the design simulation and tests.

ALFE2 has 128 register bits (16 bytes) for initial configuration, such as adjustable timing constant, baseline voltage, and input impedance, via an I²C interface. The I$^2$C target functional block, including the registers, is protected using a triple modular redundancy (TMR) technique to enhance resilience against single-event effects (SEEs). The TMR design supports refreshment, where an error bit that disagrees with the other two bits in a triplicated cell is set to the value of the majority (the other two bits) in less than one nanosecond.

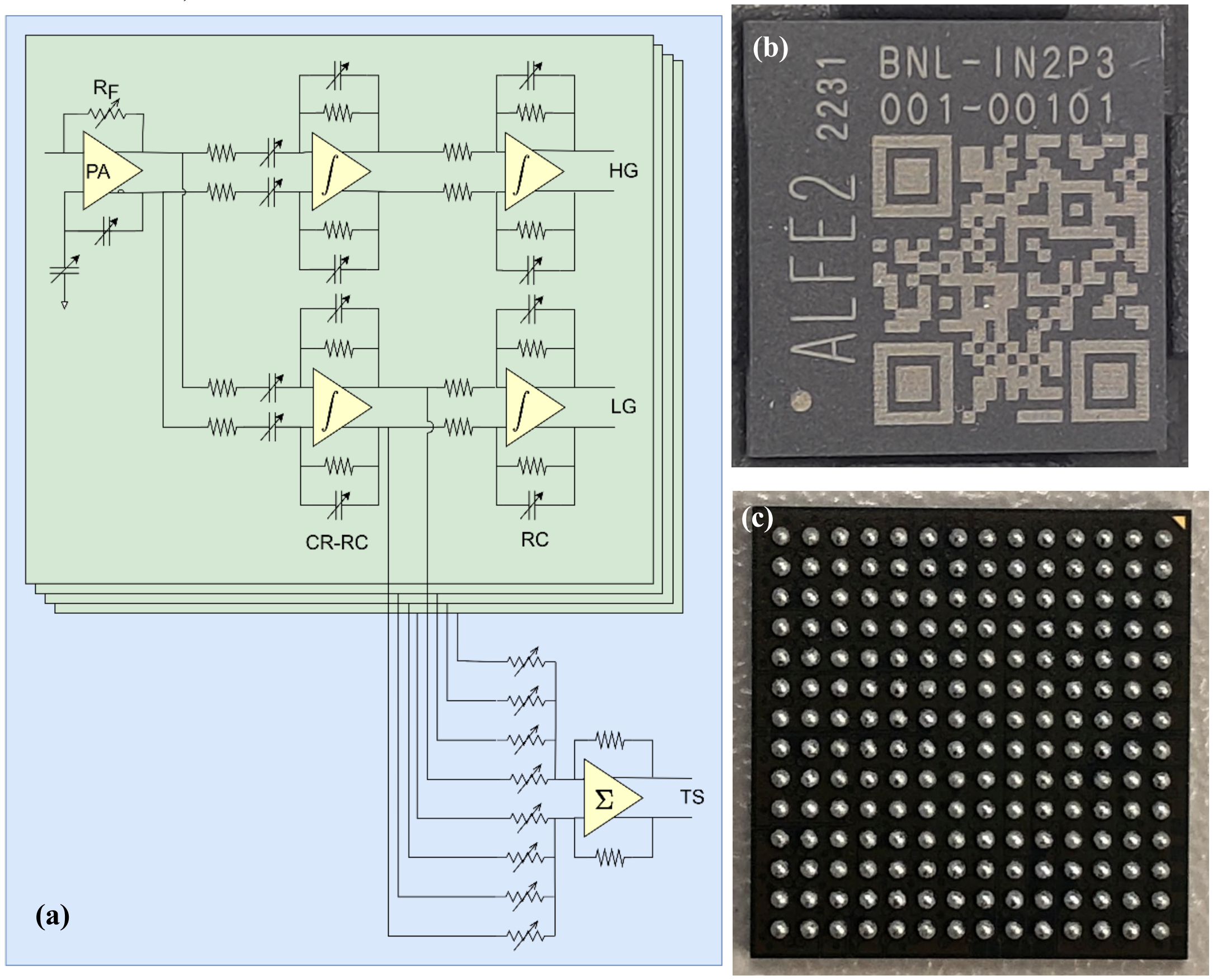

**Figure 1**. Block diagram (a), top-view photograph (b), and bottom-view photograph (c) of an ALFE2 chip.

ALFE2 is manufactured in a 130 nm CMOS process. ALFE2 operates with two power domains (1.2 V and 2.5 V). The silicon die (4.55 mm × 5.00 mm) is packaged in a ball grid array (BGA) with dimensions of 12 mm × 12 mm × 1.473 mm. ALFE2 has 196 balls with a ball pitch of 0.8 mm. The top-view and bottom-view photographs of an ALFE2 chip are shown in Figure 1(b–c).

Slightly over 80,000 chips were manufactured and packaged from 46 wafers. Approximately 50,000 ALFE2 chips are required to populate the entire ATLAS LAr Calorimeter front-end boards [10] (including spares and a safety margin). A yield of 70% is therefore required.

Over 10% of the ALFE2 chips were tested to evaluate both the chip performance and the quality-control (QC) test system. Irradiation tests, including total ionizing dose (TID) and SEE tests, were conducted to evaluate radiation tolerance as a quality assurance (QA) procedure. The setups and results of these QC and QA tests are presented in this paper.

## 2. QC test setup and preliminary results

### 2.1 QC test setup

The ALFE2 chips are tested at Brookhaven National Laboratory (BNL) and the Laboratory of the Physics of the two Infinities Irène Joliot-Curie (IJCLab). The QC test system consists of two primary parts: (1) the core setup, used for characterization, irradiation, and QC tests, and (2) a robotic arm and a controller for the automated QC test of ALFE2.

#### 2.1.1 Core QC setup

The block diagram and a photograph of the core setup [11][12] are shown in Figure 2. This setup comprises an ALFE2 test board, a front-end test board (FETB), and a calibration pulser board.

The ALFE2 test board hosts a socket to accommodate the ALFE2 chip under test. Three socket types (VA Innovation Part No. VAT-B196PN-2060, Knight Auto Part No. 12x12 mm 196 BGA 0.80 mm pitch test socket, and Ironwood Electronics Part No. CBT-BGA-6026) have been tested. The first type is an open-top socket for the robotic test system and irradiation tests, while the other two are clamshell sockets used only for characterization.

The ALFE2 test board is connected to the calibration pulser board using a 50 Ω (Z shown in the figure) coaxial cable to approximate the detector connections. The ALFE2 test board integrates radio-frequency (RF) relays (Omron Electronics Part No. G6K-2P-RF) to select between two detector loads (Cd) in different input-impedance configurations (330 pF for 50 Ω or 1.5 nF for 25 Ω). In addition to load selection, the test board fans out a single calibration pulse to the four input channels of ALFE2. Resistors Rinj and Ra ensure correct 50 Ω termination.

FETB is a versatile data acquisition system that integrates analog circuits and a field-programmable gate array (FPGA). It features two 8-channel, 16-bit ADCs (Texas Instruments Part No. ADS52J65) with 13-bit ENOB and a maximum rate of 125 MS/s (a 40 MS/s rate is used in the test). FETB includes a system-on-module (SoM) (Enclustra Mercury+ XU1 series). This SoM is based on a Xilinx UltraScale+ Zynq multi-processor system-on-chip (MPSoC) FPGA. The embedded ARM processors on FETB run a lightweight operating system (PetaLinux) for real-time data acquisition and analysis. FETB configures the ALFE2 chip under test. Major programs are written in Python, while performance-critical tasks are implemented in C++. It takes approximately 22 seconds to measure the performance of a single chip on the core setup. For comparison, the average test time is 1.5–2 minutes for a robotic test.

The ALFE2 test board can be mounted onto FETB as a daughterboard via an FPGA mezzanine card (FMC) connector. An FMC extension cable can also be inserted between the two boards for irradiation tests.

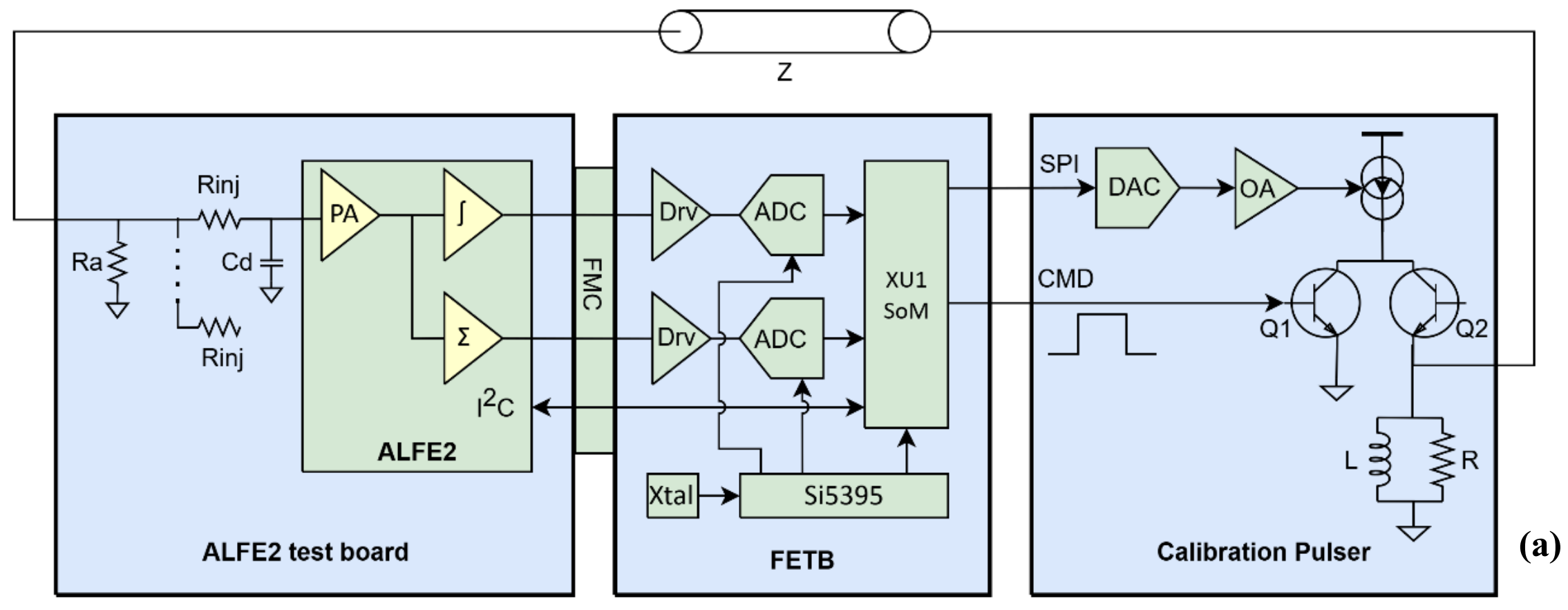

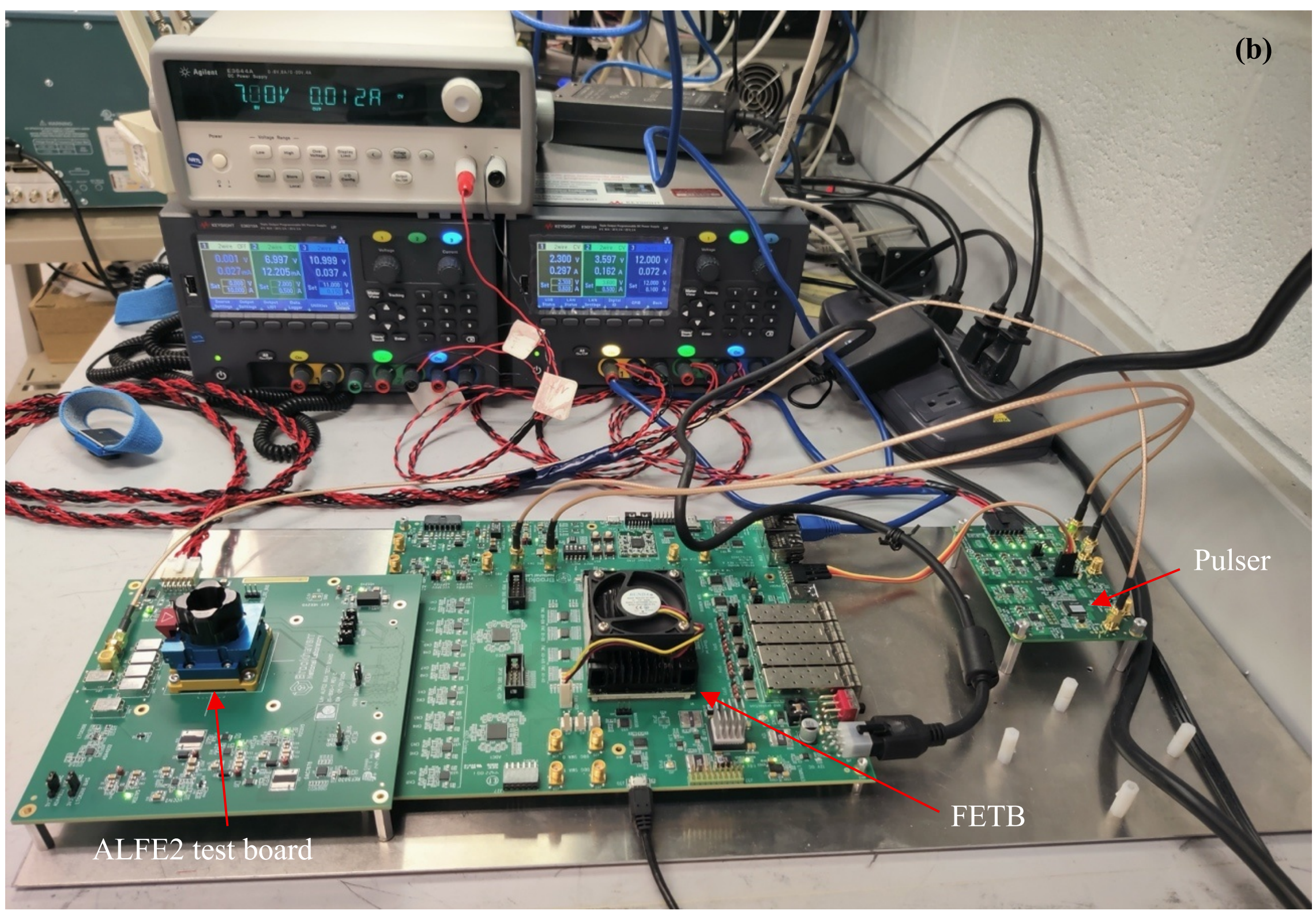

**Figure 2.** Block diagram (a) and photograph (b) of the core test setup.

The calibration pulser board injects a precise current signal into the test board. The core circuit consists of a precise DC source, a differential pair of transistors (Q1 and Q2), an inductor (L), and a resistor (R). By default, Q1 is turned off, and Q2 is turned on, allowing the current to flow through the inductor. A positive calibration command (CMD) pulse from FETB turns on Q1 and turns off Q2. Subsequently, the magnetic energy stored in the inductor decays through the resistor and the coaxial cable connected to the ALFE2 chip, producing an exponential decay pulse. The current source can be adjusted via a 16-bit digital-to-analog converter (DAC) (TI Part No. DAC8830). The DAC is configured using a serial peripheral interface (SPI). FETB configures the DAC and provides the calibration command pulse required. The calibration pulser board utilizes a low-offset operational amplifier (OA) ASIC. The ASIC was designed by IN2P3/OMEGA and employed on the existing ATLAS LAr calibration board [13]. It is noted that the current source and Q1 are integrated as parts of the amplifier ASIC.

### 2.1.2 Robotic QC test setup

In addition to the core test setup, the automatic QC test system incorporates a robotic arm controlled by a robot controller. Before testing, the robotic arm picks up the chip from a tray and places it into the socket. After testing, the arm retrieves the chip from the socket and sorts it into pass/fail trays based on the test results and grading criteria. During the evaluation of the QC test, however, the tested chips were placed back onto their original tray before the grading criteria were finalized. Photographs of the robotic test system at BNL and IJCLab are shown in Figure 3. The system is installed on a 5 ft × 10 ft optical table. The robotic arm is a 4-axis step-motor system (IGUS MOT-AN-S-060-035-060-M-C-AAAC for the X-axis and Y-axis, IGUS MOT-AN-S-060-035-060-M-D-AAAD for the Z-axis, and SMC VT307-5D01-02F-Q for the Z'-axis). The first two axes (X and Y) are horizontal. The Z-axis opens or closes the socket by engaging or disengaging the shoulder of the open-top socket. A vacuum suction cup is mounted on the robotic arm, which independently moves up and down as the fourth (Z') axis. The suction cup is connected to a 20-liter gas reservoir (Festo Part No. 534845), which is evacuated by a vacuum pump (KNF Model UN816.1.2KT.18). The vacuum pump can be turned on or off via a remote-access power distribution unit (APC Part No. AP7900B). The reservoir allows the vacuum pump to be cycled less frequently. The robotic arm is also equipped with a camera (Somikon Model No. PX-3714-675). The camera reads QR codes etched on the chip surfaces for chip identification, printed on the optical table for the system origin, or attached to the test board for socket position reference. The robot control board houses motor drivers and two Arduino modules: an Arduino Mega 2560 module controls the step motors, and an Arduino Uno module toggles the vacuum valve. The setups at IJCLab and BNL are essentially identical except that the BNL robot is slightly wider and longer than the IJCLab unit.

## 2.2 QC evaluation results

6,531 chips at IJCLab and 3,197 chips at BNL have been tested by September 2025, accounting for over 10% of the total chip inventory. A detailed analysis of these results was performed. Over 97% of the tested chips are functioning and meet all specifications. This is higher than the 70% yield required to populate all the front-end boards of the ATLAS Liquid Argon Calorimeter. To ensure optimal performance and minimize the risk of delivering dysfunctional chips, a performance grading system was introduced. Grade A (fully qualified/good) is defined for chips whose performance is entirely within the grading criteria, which are stricter than the specifications. Grade B (marginally qualified/average) is defined for chips that exhibit one or more marginal failures from the grading criteria, excluding power dissipation and $I^2C$ communication performance. Grade F (failed/bad) is defined specifically for chips with marginal power dissipation or $I^2C$ communication issues. Power dissipation and $I^2C$ performance are treated separately because they are less sensitive to socket performance degradation (discussed later) than other analog performances. Grade-F chips will not be populated on front-end boards. Grade-B chips will not be initially delivered and may be retested if needed. It is suspected that most of the Grade-B chips are linked to the imperfect connectivity of a single output channel in the socket and are therefore recoverable. The grading criteria are determined through offline analysis. The major specifications and corresponding grading criteria are summarized in **Table 1**. The yield (i.e., the percentage of the Grade-A chips) exceeds 86.7% at IJCLab and 88.1% at BNL, respectively.

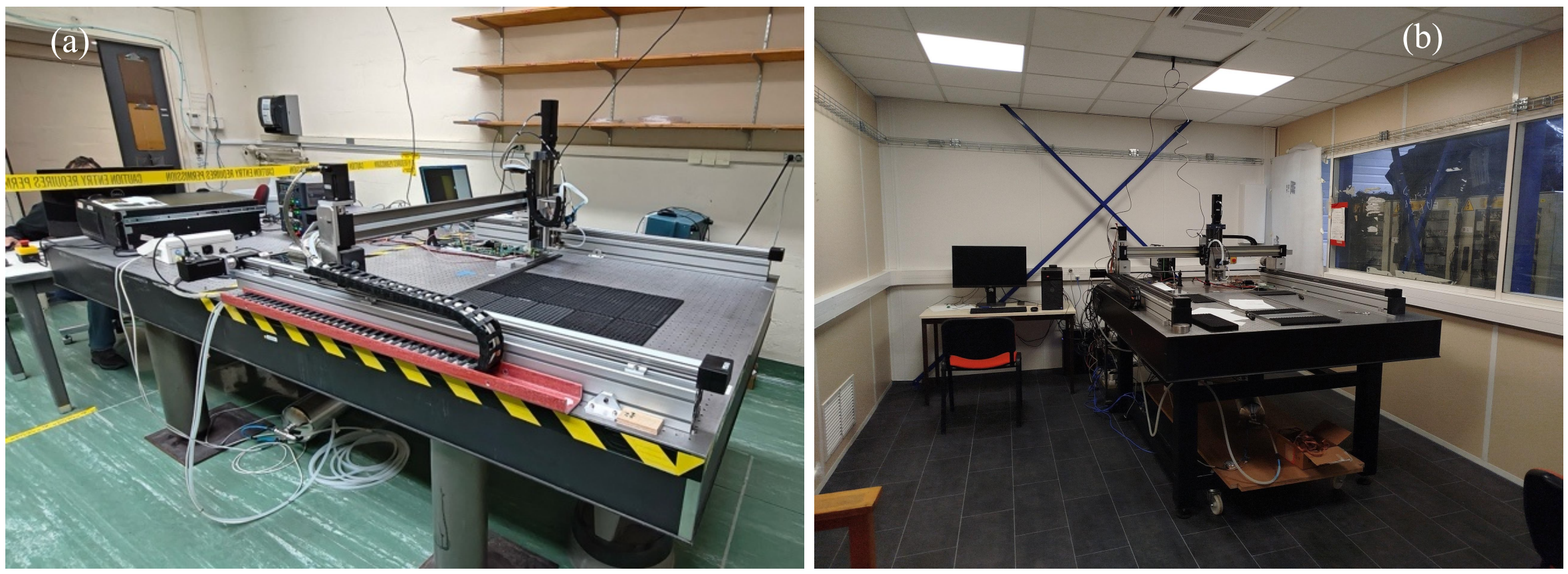

**Figure 3**. Photographs of the robotic test system at BNL (a) and IJCLab (b).

**Table 1**. Specifications and grading criteria of ALFE2.

| Parameters | | Specifications | Grading criteria |
|---|---|---|---|
| Max input current (mA) | | 0.08 (HG 50 Ω), 0.45 (HG 25 Ω), 2 (LG 50 Ω), 10 (LG 25 Ω) | |
| Nominal load (pF) | | 330 pF (50 Ω), 1500 pF (25 Ω) | |
| Gain ratio (HG/LG) | | 22±5 | 23.1–23.9 (50 Ω), 23.0–23.8 (25 Ω) |
| Peaking time (ns) | HG&LG 50 Ω | 38±5 | 36–40 |
| | HG&LG 25 Ω | 46±5 | 44–48 |
| Uniformity | Baseline | N/A | <5% (HG), <3% (LG) |
| | Gain | N/A | <2% (50 Ω), <1.5% (25 Ω) |
| | Peaking time | N/A | <2% (HG 50 Ω), <1.5% (HG 25 Ω, LG) |
| INL | HG | <0.2% | <0.17% (50 Ω), <0.15% (25 Ω) |
| | LG 80% range | <0.5% | N/A |
| | LG full range | <5% | <0.25% (50 Ω), <0.4% (25 Ω) |
| ENI (nA) | HG 50 Ω | <120 | <60 |
| | HG 25 Ω | <350 | <200 |
| Power dissipation | | <650 mW | 265–305 mA (1.2 V), 95–108 mA (2.5 V) |
| $I^2C$ phase margin | | N/A | >250° |
| Yield | | >70% | >85% |

The integral non-linearity (INL) is the most critical performance metric for ALFE2. The INL is measured by injecting a series of signal pulses with increasing amplitudes. The INL histograms of the LG outputs in the 50 Ω configuration from the two test sites are compared in Figure 4. For the LG outputs, ALFE2 performed far better than the specification of 5% in the full dynamic range and 0.5% in the 80% range. The grading criteria for the LG outputs are set at 0.25% in the 50 Ω configuration and 0.40% in the 25 Ω configuration in the full range, respectively. For the HG outputs, the grading criteria are 0.17% (50 Ω) and 0.15% (25 Ω), respectively, compared with the specification of 0.2%.

The equivalent noise current (ENI) performance is crucial for ALFE2 to achieve the desired INL. The ENI is calculated as the measured root-mean-square (RMS) noise divided by the corresponding gain. The ENI histograms of the HG outputs in the 50 Ω configuration from the two test sites are shown in Figure 5. The measured noise on the robotic test stands is slightly higher than that on the standalone core test setup because ALFE2 picks up additional noise radiated by the robot. The ENI of ALFE2 is superior to the specifications of 120 nA (50 Ω) and 350 nA (25 Ω). Therefore, the relatively loose grading criteria of 60 nA (50 Ω) and 200 nA (25 Ω), which fall beyond the x-axis range of the histogram plots, are deemed sufficient.

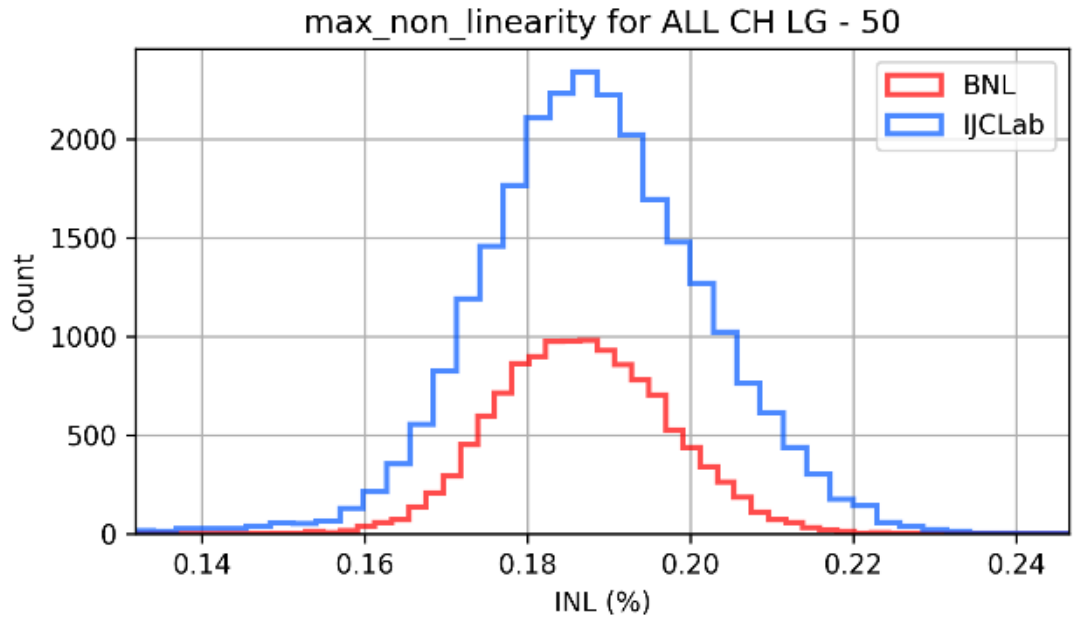

**Figure 4**. INL histograms of the LG outputs in the 50 Ω configuration.

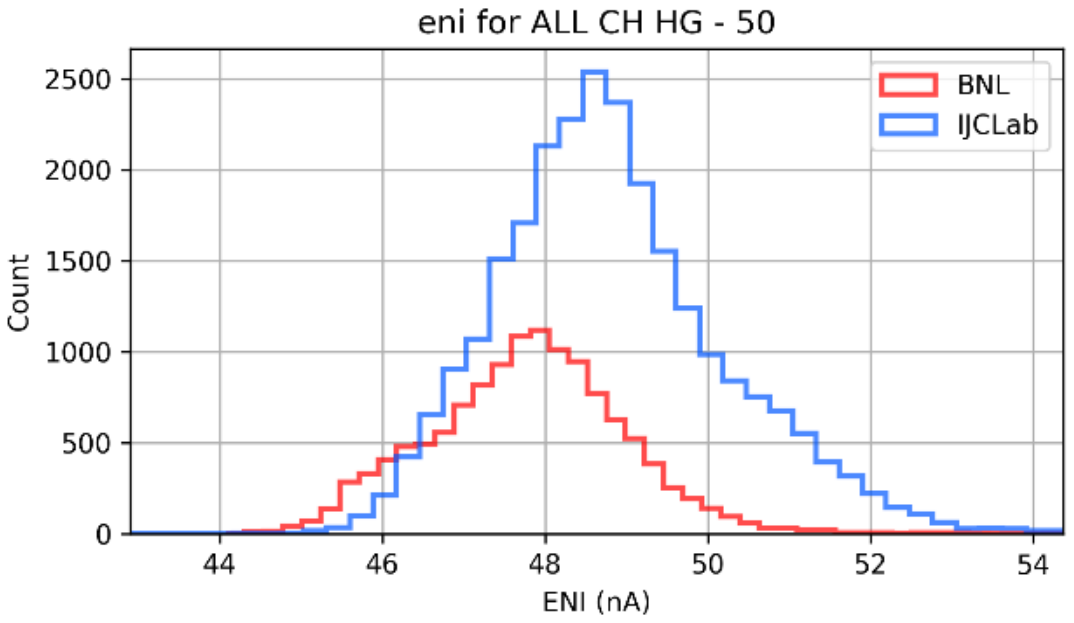

**Figure 5**. ENI histograms of the HG outputs in the 50 Ω configuration.

As individual gains will be calibrated in the ATLAS application, there is no precise specification for the absolute gain. However, the gain ratio remains important for ALFE2. The histograms of the gain ratio in the 25 Ω configuration between the two sites are shown in Figure 6. The measured gain ratio distribution is tight, so the grading criteria are set to 23.1–23.9 (50 Ω) and 23.0–23.8 (25 Ω), compared with the specified range of 22±5. A noticeable yet only about 1% difference is observed between the two sites. This is likely due to variations in the load capacitance (tolerance 1%) on the ALFE2 test board and inductance (tolerance 5%) on the calibration pulser. The planned future exchange of a full tray between the two sites could further refine these comparisons.

The histograms of power dissipation in the 25 Ω configuration are shown in Figure 7. The grading criteria for the supply currents are set at 265–305 mA for 1.2 V and 95–108 mA for 2.5 V, respectively. Correspondingly, the maximum criterion for the total power consumption is 636 mW, which is below the 650 mW specification (black vertical dash line shown in the figure).

During the QC test, socket performance was observed to degrade with the number of chip insertions. A magic eraser (VA Innovation Part No. VAC-BS-0707-090), a nylon brush, and a compressed-nitrogen blower were used to recover socket performance [14]. Since the corners of the socket are difficult to access, a sharpened stick cut from the magic eraser was held with a pair of tweezers to clean the pogo pins at the four corners of the socket. After socket cleaning, a handheld USB microscope (AmScope model No. 80200-P) was utilized to inspect the socket.

To determine the required cleaning frequency, the cumulative counts of failed chips and marginally qualified chips versus the cumulative number of chips tested since the last socket cleaning are shown in Figure 8(a). Following the completion of the first tray (189 chips), a modest uptick in the cumulative number of outlier chips is observed, which escalates as the second tray nears its end.

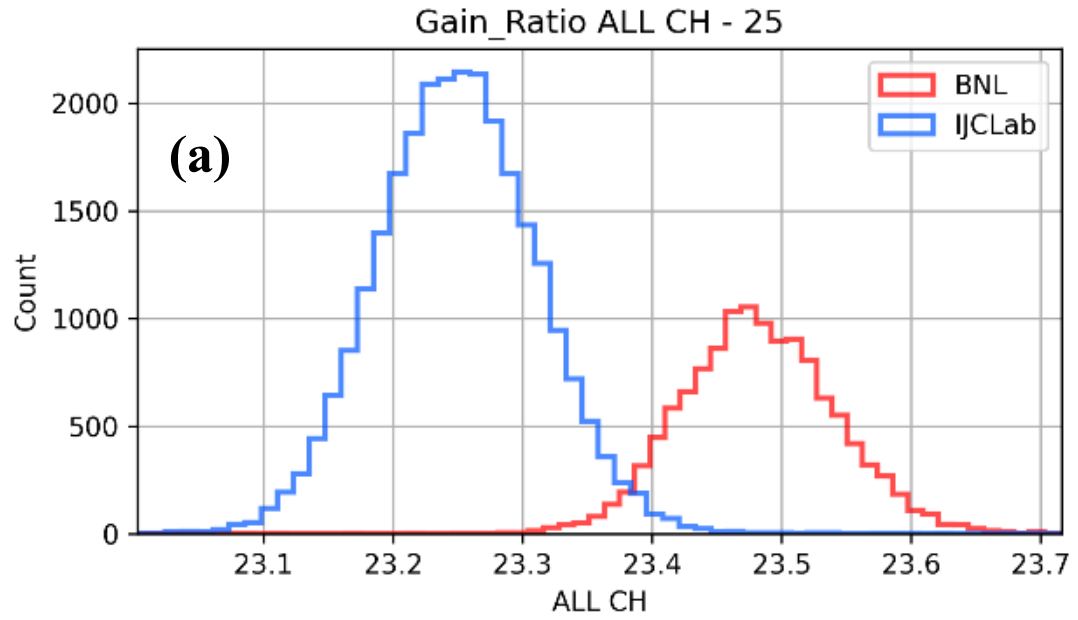

**Figure 6**. Histograms of the gain ratio in the 25 Ω configuration.

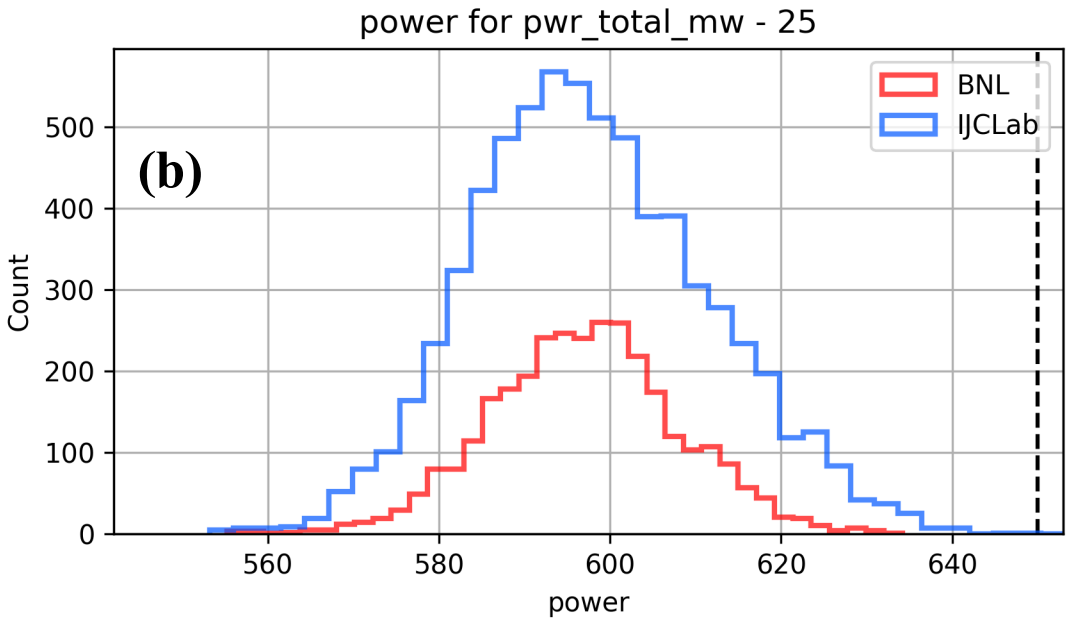

**Figure 7**. Histograms of the power dissipation in the 25 Ω configuration.

Uniformity, as well as gain ratio, is a useful metric for determining socket performance degradation. Uniformity is defined as the (maximum - minimum) / average value across the four channels for the baseline, gain, and 5%–100% peaking time. A lower uniformity value indicates better performance. One of the most sensitive parameters is the baseline uniformity. The series of the baseline uniformity of the HG outputs in the 50 Ω configuration versus the cumulative number of chips tested since the last socket cleaning is shown in Figure 8(b). The baseline uniformity follows the trend seen in the accumulation of failed and marginally qualified chips. The grading criteria for the baseline uniformity are set at 5% for the HG outputs and 3% for the LG outputs. The grading criteria for the other uniformities are listed in Table 1.

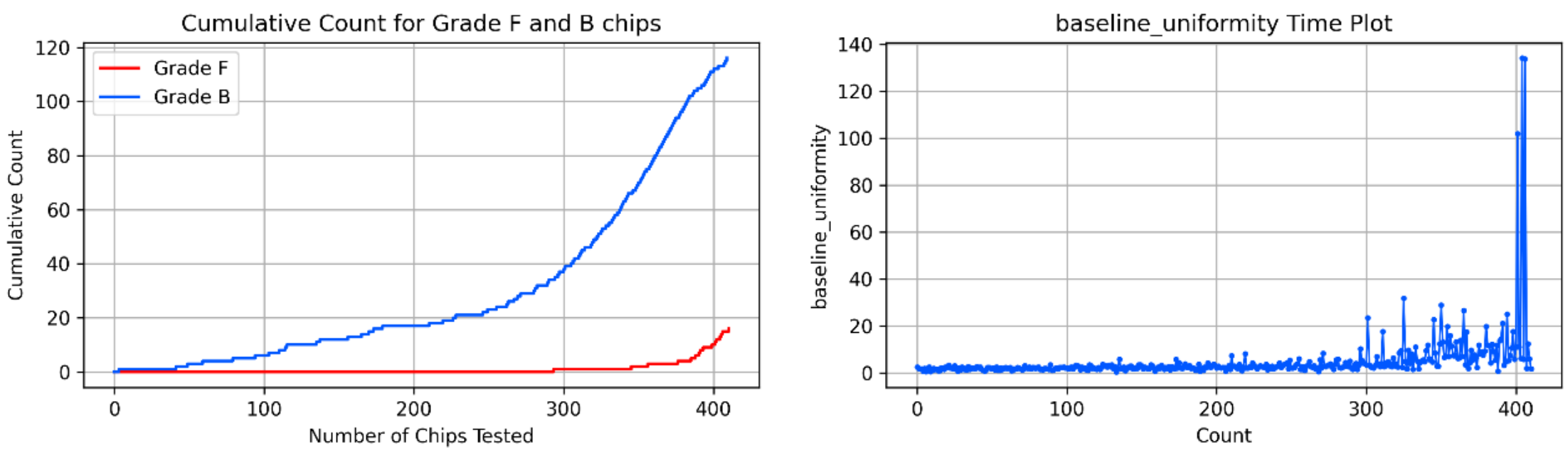

Figure **8**. Cumulative count of Grade-B and Grade-F chips (a) and baseline uniformity of the HG outputs in the 50 Ω configuration (b) versus the cumulative number of chips tested since the last socket cleaning.

From these findings, the socket will be cleaned after testing each tray during the formal QC test. Based on our current experience, each socket maintained its performance for over 6,000 insertions with routine cleaning.

## 3. Irradiation tests

ALFE2 is designed to operate in a harsh radiation environment and must meet stringent radiation tolerance requirements [15][16]. The TID specifications are 1.40 kGy (referenced to silicon dioxide, not indicated hereafter) for the front-end boards in the barrel area and 0.21 kGy for those in the end-cap area. The SEE specifications are $1.0 \times 10^{13}$ hadrons /cm$^2$ (with energy > 20 MeV) for the barrel front-end boards and $1.2 \times 10^{12}$ h/cm$^2$ for the end-cap front-end boards. Safety factors of 1.5 (TID) and 3 (SEE) are included in these values, respectively. These specifications are determined to cover a fluence of 4000 fb$^{-1}$ over 10 years of HL-LHC operation. To ensure ALFE2 meets the required radiation tolerance, a TID test and an SEE test were conducted. Since ALFE2 is designed in CMOS technology, it is insensitive to non-ionizing energy loss (NIEL) effects, and thus no dedicated NIEL test is required.

### 3.1 TID test setup and results

The TID test was performed from August to November 2024 at BNL. Thirty chips (from one engineering wafer lot and two production wafer lots) were irradiated with $^{60}$Co gamma rays. One extra sample from each wafer lot was used as a non-irradiated reference. The dose rate was 100 Gy/hour. The total dose for each sample ranged from 2.23 kGy to 9.59 kGy. A photograph of the irradiation test setup is shown in Figure 9. During irradiation, each ALFE2 chip was placed in an open-top socket mounted on an ALFE2 test board with the socket opening facing the gamma-ray beam. The test board did not contain RF relays or load capacitance. Instead, the load capacitance was implemented on a separate board. A FETB, a calibration pulser board, and DC voltage power supplies were placed a few meters away from the test board and shielded with lead bricks. After irradiation, each sample underwent annealing for approximately 23 hours at room temperature and for 7 days in an oven at about 100 °C. During irradiation and room-temperature annealing, each sample operated in the 50 Ω configuration, with major performances (including supply currents) and board temperatures continuously monitored. During high-temperature annealing, each sample was powered using power-on-reset configurations, and only supply voltages, supply currents, and temperature were monitored. After annealing, all samples were re-evaluated using the routine QC test procedure in both the 25 Ω and 50 Ω configurations. The temperature varied within 0.2 °C during irradiation. During room-temperature annealing, the temperature fluctuated within 2 °C. For the high-temperature annealing, the temperature was set at approximately 100 °C, fluctuating within 15 °C.

The INL performance change subjected to the highest TID dose is shown in Figure 10(a). The two vertical red dashed lines indicate the start and end times of the gamma-ray beams, while the vertical blue dashed line marks the time corresponding to the specified dose of 1.4 kGy. No discernible change was observed in the INL during either irradiation or annealing. The INL was higher during irradiation than in the QC test because of the one-meter FMC cable connecting the ALFE2 test board with the FETB. The ENI performance change during irradiation is shown in Figure 10(b). Fluctuations in noise from which the ENI was converted were observed to be about 5 μV for the HG and LG outputs, about 15 μV for the TS output (gain of one), and about 30 μV for the TS output (with a gain of three). However, these fluctuations were comparable to a least significant bit (LSB) (30.5 μV) of the ADCs used on the FETB and less than that (61.0 μV) of the ADC ASIC [17] planned for the ATLAS LAr Calorimeter, indicating that the noise was not materially affected. The noise remained stable during both room-temperature and high-temperature annealing.

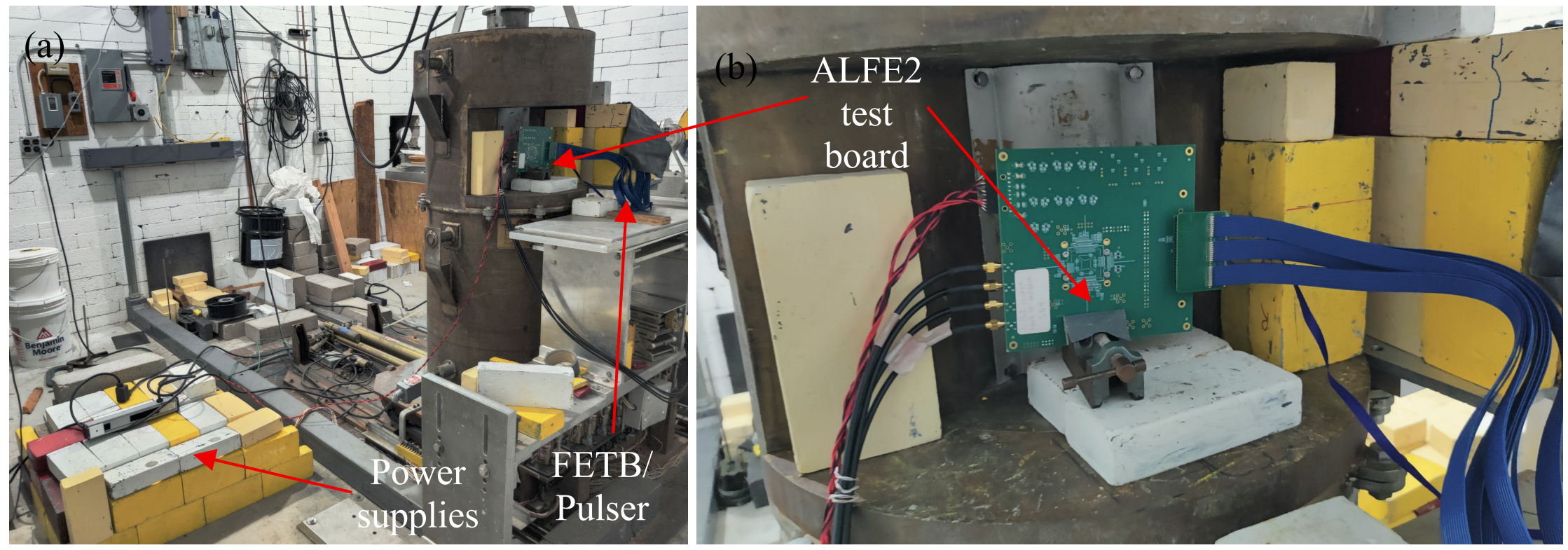

**Figure 9**. Photographs of the overall TID test setup (a) and the test board located at the beam exit (b).

The 1.2 V and 2.5 V supply currents remained stable during irradiation, room-temperature annealing, and high-temperature annealing phases. The changes observed in baseline and gain were within ±0.5% and ±0.1%, respectively. No discernible change was observed in the 5%–100% peaking time. All samples passed the routine QC test after irradiation and annealing. These test results confirm that the analog performance of ALFE2 meets the TID radiation-tolerance requirements.

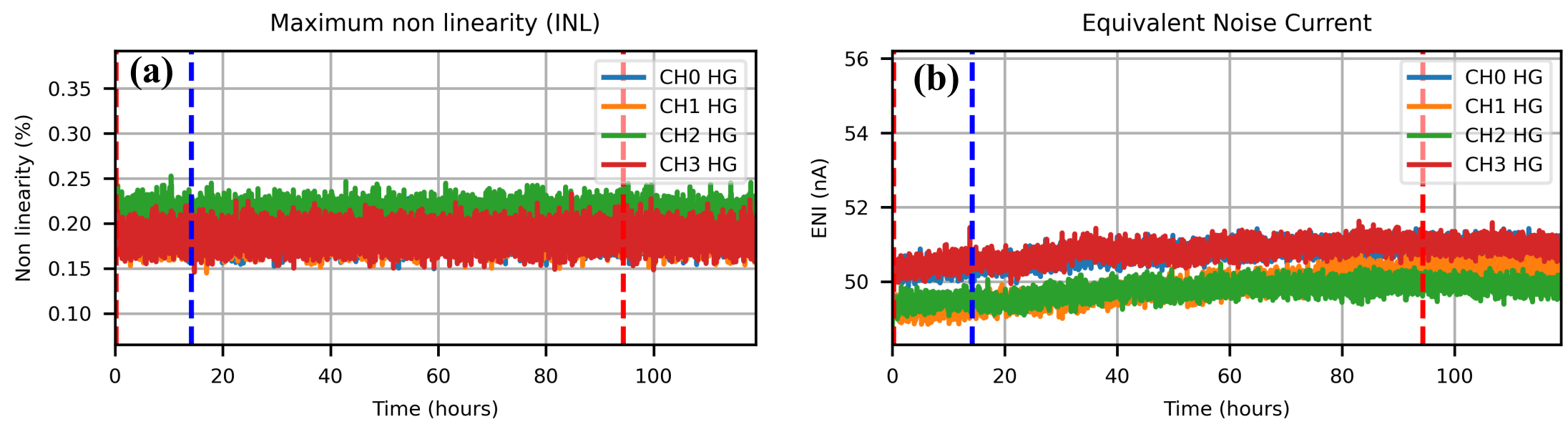

**Figure 10**. INL (a) and ENI (b) of the HG outputs during irradiation.

Measurable changes were observed in the $I^2C$ performance during the TID test. The design of ALFE2's $I^2C$ target functional block, which was inherited from a previous project [18][19][20], requires an external 40 MHz system clock. Furthermore, this clock must be maintained at a specific phase relative to the $I^2C$ signals (SCL and SDA). The SCL signal supports both the open-drain mode and the CMOS driving mode. In CMOS mode, the SCL signal is driven to the high level with a CMOS gate, whereas in open-drain mode, the SCL signal is pulled up to the high level with a resistor [21].

It is important to understand the relationship between the $I^2C$ error rate and the relative system clock phase and subsequently select an optimal clock phase. For this purpose, the system clock phase was swept from 0° to 360° (corresponding to a clock period of 25 ns) with a step of 1.5° at the $I^2C$ clock frequencies of 100 kHz, 200 kHz, 400 kHz, and 1 MHz. At each phase, all the registers of ALFE2 were written and read 10,000 times. All register operation exceptions (e.g., timeout and no acknowledgement) and data mismatch (i.e., the read values not matching the written values) were counted as errors. A typical relationship between the error counts and the system clock phase for the QC setup and the TID test setup is shown in Figure 11. The CMOS mode is used in the QC test, while the open-drain mode is used in the TID test. The 1 MHz operating frequency was unreliable in the open-drain mode. As can be seen in the figure, there are no errors except for two phase lobes, one of which is substantially higher than the other. The low and high lobes likely correspond to the coincidence of the transitive edges of the SDA signal and the rising edges of the SCL signal, respectively, with the system clock. The transitive edges of SDA are internally sampled only during the START and STOP conditions of an I$^2$C transaction, whereas the SCL signal is sampled at each of its rising edges. Both error lobes are noticeably wider in the open-drain mode than in the CMOS mode. No discernible differences were observed between different frequencies in the CMOS mode, whereas in the open-drain mode, the lobes markedly varied from one frequency to another. The difference between the two modes is likely attributed to the fact that the rising edges of SCL are much slower in open-drain mode than in CMOS mode. The presence of the FMC cable in the TID test setup makes the lobes even wider and their frequency-dependent variations more pronounced.

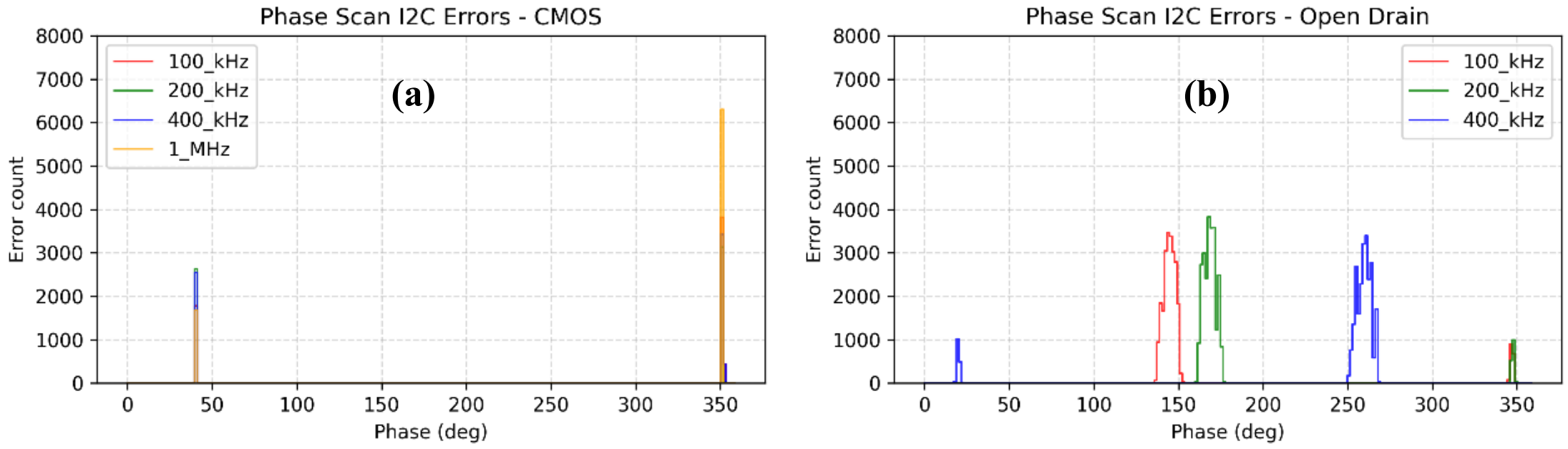

**Figure 11**. Error count versus the clock phase for the CMOS (a) and open-drain (b) modes.

The I$^2$C target functional block must operate when the clock phase falls within either of the two error-free gaps between the two lobes. There are two such gaps, with one being substantially wider than the other. The width of the wider error-free gap is defined as the phase margin, and the center of this wider gap is selected as the optimal phase. The histogram of the phase margin at the operational frequency of 1 MHz is shown in Figure 12. The grading criteria are set for a minimum phase margin of 250°. In the QC test, all registers of ALFE2 are written and read only 64 times at 400 kHz and 1 MHz to save time. After the optimal phase is determined, the registers are tested at that optimal phase 1,000 times to verify error-free operation in the QC test.

A typical change in I$^2$C phase margin at 400 kHz during the TID test is shown in Figure 13. The phase margin shrank by less than 40°. The phase margin in the CMOS mode (~310°) is markedly larger than in the open-drain mode (~200°). Therefore, the observed 40° difference in the phase margin is not expected to impact the ATLAS application.

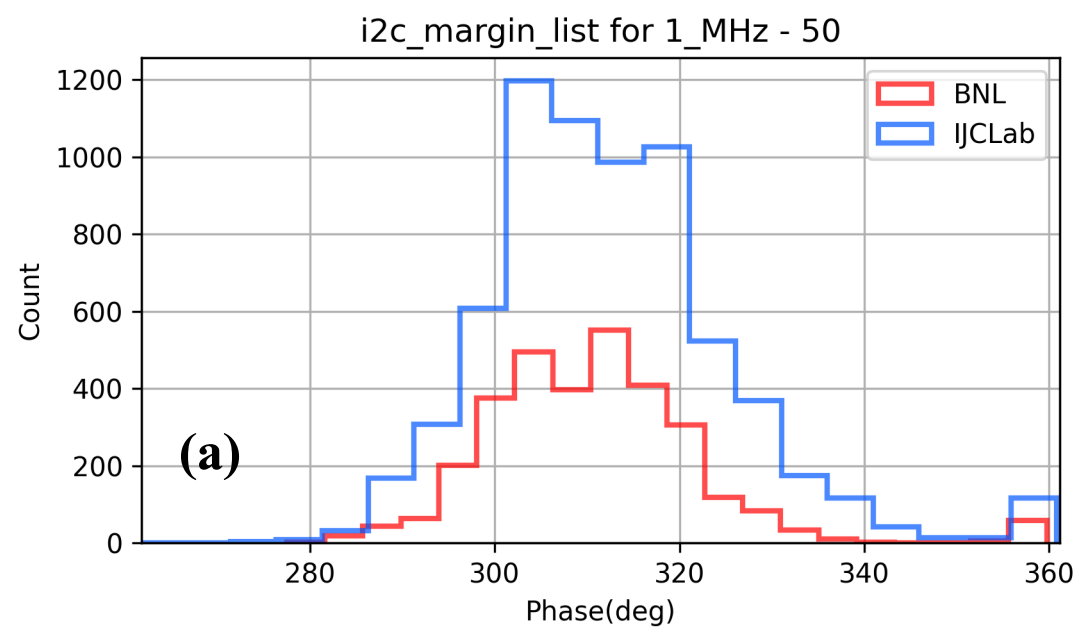

**Figure 12**. Histogram of the $I^2C$ phase margin at 1 MHz in the QC evaluation.

**Figure 13**. Changes in the $I^2C$ phase margin at 400 kHz during irradiation.

### 3.2 SEE test setup and results

The SEE test was conducted on April 22–23, 2023, at Massachusetts General Hospital. A 226 MeV proton beam with a profile diameter of 1.5 cm and a flux ranging from $7\times10^9$ to $3.5\times10^{10}$ p/cm$^2$/s was utilized. A photograph of the test setup is shown in Figure 14. An ALFE2 chip was placed in an open-top socket, aligned with the beam such that the socket opening faced the beam exit. The FETB was placed two meters away from the test board and shielded with lead bricks. All power supply units were located six meters away and were similarly shielded.

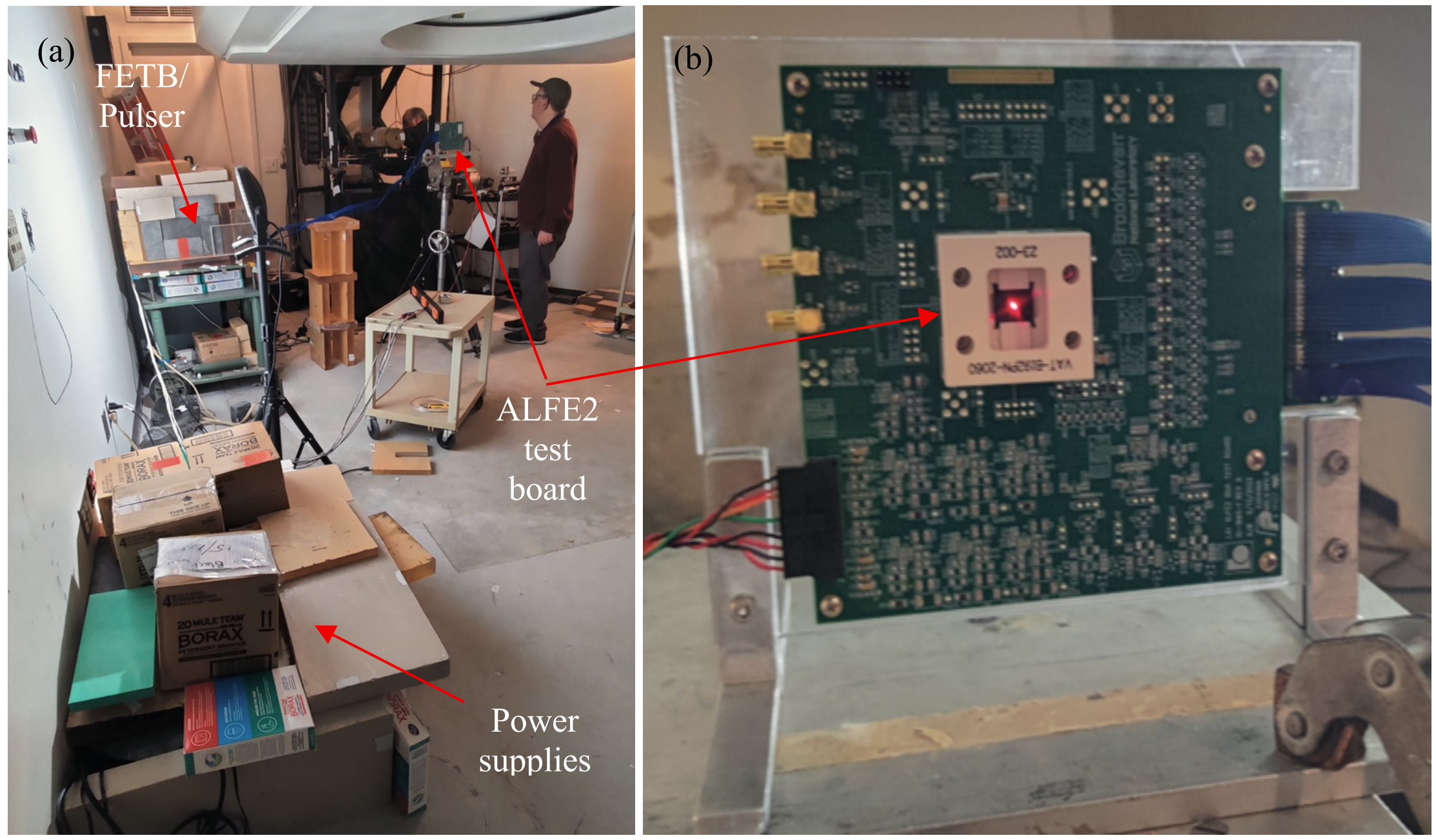

**Figure 14**. Photographs of the SEE test setup.

The internal registers of ALFE2 can be accessed in two modes: single-byte access or multiple-byte block access. The multiple-byte block mode, which will be used for the future ATLAS application, was adopted for this test. During irradiation, the registers were repeatedly written and read to check for SEU occurrences. Ninety-three register bits (out of 128 total) were first written with incremental values in the sequential block mode at 400 kHz. The remaining bits were kept at default values to maintain the ASIC's functionality.

Three chips were tested without refreshment. A total of 108 single-event upsets (SEUs) were observed across these three chips, corresponding to a total fluence of $2.74\times10^{14}$ protons/cm$^2$. By combining the results of the three chips, the SEU cross-section is estimated to be $3.56\times10^{-15}$ cm$^2$/bit at the 95% confidence level.

Four chips were irradiated with refreshment enabled. No SEU errors were detected, confirming the effectiveness of the refreshment mechanism. By combining the results of the four chips, the upper limit of the cross-section was estimated to be $1.35\times10^{-16}$ cm$^2$/bit at the 95% confidence level. This translates to an extrapolated, acceptable upper limit error rate of 4.6 errors per day for the entire HL-LHC ATLAS LAr Calorimeter at the 95% confidence level. The error rate estimation accounts for the fluence difference between the barrel area (accommodating 896 front-end boards) and the end-cap area (accommodating 628 front-end boards).

All seven chips survived exposure to ionizing doses that were higher than (six chips) or close to (the remaining chip) the TID specification. All seven chips remained functional after the SEE test, with no significant performance degradation observed.

## 4. Conclusion

A robotic system has been developed for the QC test of ALFE2. Over 6,000 chips and 3,000 chips have been evaluated at IJCLab and BNL, respectively. The evaluation allowed us to establish the grading criteria. Using these criteria, a yield of over 85% was achieved in the evaluation tests, and these criteria are now being applied to the ongoing full-production QC. Irradiation tests were performed for the QA of ALFE2. No significant performance degradation was observed during the TID test. Based on the SEE test results, an acceptable error rate of fewer than 4.6 SEUs per day is extrapolated for the entire ATLAS LAr Calorimeter during HL-LHC operation.

## Acknowledgments

We are deeply grateful to everyone whose efforts contributed to the design and testing of ALFE2. Our appreciation extends first to G. de Geronimo for his foundational design work on HLC1, the predecessor of ALFE. We are also thankful to the IN2P3/OMEGA team, including S. Blin, S. Conforti, C. de la Taille, M. Berni, and N. Seguin-Moreau, for their valuable collaboration. We further acknowledge the support of V. Manthena and S. Miryala from the Instrumentation Department at BNL, as well as S. Michelis from CERN. We also extend our thanks to T. Cotroneo, S. Patra, R. Islam, Z. Kekula, and M. Lowy for their contributions to quality control testing during their internships at BNL. Finally, we are grateful to E. Cascio from the Proton Therapy Center at Massachusetts General Hospital in Boston for his support during the SEE test.